\documentclass{PoS}
\usepackage{graphicx}
\usepackage[]{epsfig}
\usepackage{epstopdf}
\input{epsf.sty}

\usepackage{texdraw}
\newcommand {\bi} {\bibitem}
\newcommand {\be} {\begin{equation}}
\newcommand {\ee} {\end{equation}}
\newcommand {\bea} {\begin{eqnarray} }
\newcommand {\eea} {\nonumber \end{eqnarray}}
\newcommand {\eps} {\epsilon}

\newcommand {\ba} {\overline}
\newcommand {\lan} {\langle}
\newcommand {\ran} {\rangle}

\newcommand {\cN}  {{\cal N}}

\newcommand {\bc} {\begin{center}}
\newcommand {\ec} {\end{center}}
\newcommand {\bd}{\begin{displaymath}}
\newcommand {\ed}{\end{displaymath}}

\newcommand {\Tr} {\mbox{Tr}}

\def \form#1 {eq. (\ref{#1}) }
\def \parziale#1#2  {{\partial {#1} \over \partial {#2}}}
\def \bi#1 {\typeout{#1} \item}

\newcommand{\Cblu}[1]{#1}

\newcommand{\Cred}[1]{#1}

\newcommand{\nuova}{                         }

\title{Field theory and the physics of disordered systems}

\ShortTitle{Field theory for disordered systems}

\author{\speaker{Giorgio Parisi }\\
       Dipartimento di Fisica, Sapienza Universit\`a di
  Roma,\\ INFN, Sezione di Roma I,  IPFC - CNR, P.le Aldo Moro 2, I-00185 Roma, Italy\\
        E-mail: \email{giorgio.parisi@roma1.infn.it}}

\abstract{In this talk I will present some of the main difficulties we encounter in studying the large scale behavior of disordered systems. This presentation will be done using a field theory language. The  difficulties in applying the standard renormalization group approach are due to the presence of  strong non-perturbative effects that  we do not know how to master. These difficulties are particular acute in the case of ferromagnets in random field, localized electrons, growth models and spin glasses. I will review the situation presenting at the end a promising approach based on {\sl fat} diagrams. }

\FullConference{Quarks, Strings and the Cosmos - H\`ector Rubinstein Memorial Symposium\\
		August 09-11, 2010\\
		AlbaNova (Stockholm) Sweden}

\begin{document}

\section{Introduction}
Many disordered systems are characterized by the presence of quenched disorder in the equation of motion (e.g. in the Hamiltonian). In other systems, like  structural glasses, the Hamiltonian is translational invariant and the disorder  is only present in the equilibrium configurations \cite{CAV}. Although the two classes of systems have many points in common, I will consider only systems belonging to the first class.
In disordered systems there are new phenomena that are not present for ordered systems, i.e. disordered systems display a new collective behavior. Sometimes a very  strong slowing down of the dynamics is present (more than fifteen orders of magnitude): in this case the system forms a glass.
In these thirty years there have been many progresses and breakthroughs, however in some cases there are fundamental questions that are still not answered, in spite of the fact that some of the questions have formulated twenty years ago (and may be more). 

There are many random systems whose critical properties are not understood. Technically our goal is to study the large-scale behavior of random systems using the renormalization group. This is a very interesting problem, since random systems have peculiar features (e.g. non-perturbative phenomena like Griffiths singularities) and
often conventional approaches do not work. There are many random systems where the renormalization group approach is not
developed enough.
 
Disordered systems have also  many applications beyond physics. In many case we are interested to study the collective behavior of a large heterogeneous system  of interacting agents.
There are many examples in the literature, e.g. morphology of river basins, optimization algorithms, neural networks, structure of networks in quite different contests (internet, food, finance, proteins), species dynamics, evolution\ldots

In this talk I will concentrate my attention on the unsolved problems. I have tried to construct this presentation in a way that H\`ector would have appreciated; I hope to have reached  my aim by using as far as possible the language of field theory, that  is very compact and H\`ector mastered in a magnificent way. I will be very fast on the motivations for studying the models and on the phenomenological consequences of the theories because I want to stress the root of the difficulties.

This paper is organized as follows. After the introduction in section \ref{RG} I will present some general considerations on the renormalization group; in the next section I will show which are the main difficulties in applying the renormalization group to some disordered systems. In section 4 I will discuss some open questions for  ferromagnets in a random field; in the next section I will discuss the properties of the Schr\"odinger equation in a random potential. Section 6 will be dedicated to spin glasses, in  the next section
 I will present a very interesting new approach, i.e. the topological expansion for lattice field theories, where one expands the correlations function using {\em fat} diagrams. Finally in the last section I will present some brief conclusions.

\section{The renormalization group} \label{RG}

Our aim is to understand  the collective behavior of disordered systems,  when long-range correlations are present. This may happens or at the critical point, or in a phase where a continuous symmetry is broken and therefore Goldstone Boson type excitations are present. Sometimes we are in the strange situation where the symmetry that is spontaneously broken is induced by the formalism we use (e.g. replicas).
In all these systems we want to get insight, i.e. a quantitative and qualitative understanding. In this respect the study of universal quantities (i.e. those quantities that do not depend on the detailed form of the Hamiltonian) play an important role.

As an example let us consider  an homogenous Heisenberg ferromagnet (neglecting long-range dipole interactions). For a given material, near the Curie point that separates the high temperature phase from the low temperature phase (where spontaneous magnetization is present),  the magnetic susceptibility $\chi(T)$ behaves as

\be
\chi(T)\approx {A \over (T-T_c)^{\gamma}} \ .
\ee
The exponent $\gamma$ does not depends on the material, but only the dimensions of the space (e.g. $D=3$) and it is an universal quantity.
On the contrary $A$ and $T_c$ do depend on the material.

The renormalization group is one of the most powerful tools to study the structure of the phases of a
physical system. Using the renormalization group approach one can obtain both qualitative and quantitative
results \cite{RINO}.  In this way we can construct a general theory of second order phase transitions and we can  classify the transitions into universality
classes. The values of the critical exponents can be of two different types: a) they are equal to those of
mean field theory as it happens above the upper critical dimension ($D_U$); b) they differ from those of mean field
theory (this happens below the upper critical dimension $D_U$).
 It is possible to find out when the phase transition disappears together with the low temperature
phases. The low temperature phase is absent in dimensions below the lower critical dimension ($D_L$). The exponents are usually a non-trivial function of the dimensions between $D_L$ and $D_U$. One of the great lessons of the seventies was that the study of the phase diagram of a system in the
temperature-dimension plane is very useful and it is a source of inspiration for understanding how a system
behave in our three (space) dimensional world. The study of the behavior of models in dimensions
different from 3 has been the starting point of crucial computations in three dimensions.

The successful
quantitative use of the renormalization group for ferromagnetic phase transition in 3 dimensions, started by
the recognition the 4 is the upper critical dimension (i.e. the $\phi^4$ theory has a trivial infrared stable fixed
point for positive coupling constant in 4 dimension). This observation allowed a quantitative control of the infrared stable fixed point (an consequently of the critical exponents, i.e. the anomalous dimensions of the operators), leading to the celebrated $4-\eps$ expansion for the critical exponents. Later on a similar $\epsilon$ expansion was constructed near the lower critical dimension: for
example in the case of a continuous symmetry breaking of a conventional symmetry (e.g. $O(N)$) one can derive a $2+\eps$ expansion. 

Let us consider the case of a ferromagnetic Heisenberg model, that may be represented by a conventional $\phi^4$ theory invariant under the internal $O(3)$ group (this is essentially the Goldstone model). We have already seen that the upper critical dimension is $4$ and the lower critical dimension is $2$. 
In dimensions greater than $4$ the theory exists in presence of a cutoff. In this situation infrared divergences are absent and one finds the there are no anomaly dimensions. The critical exponent $\gamma$ takes its naive mean field value, i.e. 1.
In dimensions less or equal to 2 there are no transitions.

How to compute the value of $\gamma$ at dimensions $3$? We can use or a $4-\eps$ expansion or a $2+\eps$ expansion (or also an  $1/N$ expansion at $D=3$ for a theory invariant under the $O(N)$ group). The best method is to use a fixed dimensional approach: in this case seven non-trivial terms are known (seven loops). In this way one obtains results with and errors of a few parts over $10^{-3}$.

In many systems using the renormalization group one can obtain quantitative prediction on the exponents, on the
critical equation of state and on other dimensionless quantities. One can reach these goals by using an $\epsilon$
expansion at dimensions not too far away from the upper or the lower critical dimension. If for a given system we are not yet able to compute the values of the upper and of the lower critical dimension we
miss something important in our understanding; unfortunately, as we shall see later, in many random systems we are exactly in this sad situation.

\section{Difficulties with random systems}

The behavior of strongly disordered systems is frequently very different from the behavior of pure,
homogenous systems: electron localization is a very good example of this fact. Sometimes the most dramatic effects
show up experimentally in a time domain where many different phenomena are present: very slow relaxation
and aging \cite{Aging}, memory and oblivion \cite{37} and generalizations of the usual
fluctuation dissipation relations \cite{38}.
These dynamical effects have often a static counterpart. For example in spin glasses \cite{40} the onset of very
slow relaxation and the consequent divergence of the correlation time is associated to the divergence of a
static quantity, i.e. the non-linear susceptibility \cite{41}; in spin glasses the form of generalized fluctuation
dissipation relations in the low temperature phase is related  to an other static
quantity, i.e. the probability distribution of the overlap \cite{42}. In many cases the starting point of our theoretical approach is the homogenous situation where
disorder is absent and a perturbation expansion is done in powers of the strength of the disorder. It should be
not surprising that we have serious difficulties in understanding the non-perturbative effects produced by the
disorder.

Sometimes the replica theory is used to average over the disorder. Replica theory \cite{MPV} is a wonderful tool
and it allows a very compact reformulation of the problems: often it works in a magnificent way and also it suggests new quantities to measure. However
sometimes replica theory puts the dirty under the carpet and cleaning under the carpet is not an easy job; a strong effort
must be done to decode replica theory and to give a physical meaning to the results of the replica approach. The study of
non-perturbative effects within the replica method has a lot of subtleties that are only partially understood.

Due to  these difficulties there are many physical and experimental questions on the static
behavior of random systems that do not have a well established theoretical answer.  These questions are waiting for an answer from at least twenty years, without getting an
unambiguous response. Why are we stuck in spite of all progresses that we have done in these
recent years? The problems arise from our inability to use the renormalization group for random systems in
presence of non-perturbative phenomena.
In many cases the ignorance of the upper (or lower) critical dimension forbids us to follow the standard path; this is a clear sign of lack of understanding: indeed in most of the cases this is due to the absence of a
manageable field theoretical representation that can be used as starting point. 
This is what happens for many disordered systems. I will mention some of the most important cases.

\begin{itemize}

\item Ferromagnetic Ising systems in a random field. Here we know both the upper critical dimension ($D_U=6$)
and the lower critical dimension ($D_L=2$). However dimensional reduction \cite{65} (based on supersymmetry),
that connects the critical exponents for such a system in dimension $D$ to those of a standard ferromagnet in
dimension $D-2$, is not valid at the non-perturbative level for some known reasons \cite{65}. Unfortunately the
breaking of dimensional reduction is a  non-perturbative effect and there is no consensus on how to modify the $6-\eps$ expansion in order to include this effect. Moreover there are some qualitative
features of the behavior of the model in three dimensions that are not understood (e.g. the existence of a
field with dimensions very near to zero).

\item The localization transition. Here the lower critical dimension is well known ($D_L=2$), but the
upper critical dimension is not known. In a field theory approach (where one starts from a $\phi^4$
theory with negative coupling constant) localization is due to non-perturbative phenomena and the
localization transition is related to the formation of a zero mass bound state \cite{84,83,LOC}. Different values of the upper
critical dimension have been proposed, but no consensus is reached: there are very reasonable indications
that the critical exponents are non trivial for any dimensions, but the problem should be further analyzed. 
  When
electrons interact strongly we have Mott localization and we face the same problems. New
phenomena may appear as an effect of the Mott localization. There is a good experimental evidence that in
many materials, mainly thin films, slow phenomena set in.  May be 
 the localized electrons form a kind of glass, the so-called electron
glass \cite{54,56}. Our understanding  is limited. 

\item Spin glasses at zero magnetic field. Here we know the upper critical dimension ($D_U=6$) and we
control the critical exponents in the $6-\eps$ expansion, but we do not have a good control of the infrared
behavior in the low temperature phase. There are long-range correlations (i.e. massless particles) that are
due to the existence of Goldstone modes of a symmetry that is spontaneously broken in the low temperature
phase. Unfortunately the symmetry group is in some sense an infinite dimensional group and already at zero-loops the situation is quite complex: the propagators have both poles at $k^2=0$ and at $k^2=m_i^2$, where the index $i$
goes up to infinity and the $m_i$'s accumulate toward zero \cite{57}.
We are not able do construct
the appropriate non-linear sigma-model that should be used near the lower critical dimensions. The Ward
identities of the symmetry have been identified and they have been shown to be the responsible of many
important cancellations, but we are still far from writing the final effective theory \cite{57}. 
 Usually the non-linear sigma model is obtained from the
linear model by sending to infinity the mass of the transverse modes in such a way that only the massless one
(the Goldstone modes) survive. This requires a clear cut-separation among massive and massless modes and
this is a nightmare when there is an infinite number of particles whose mass accumulates to zero.
Ironically there is an analytic computation \cite{59} of the lower critical dimension that gives the extravagant
value $D_L =2.5$. This value is confirmed by numerical simulations \cite{60} and it is based on a non-perturbative
computation of the domain wall energy: unfortunately we are unable to transport the basic ideas into the perturbative
field theory formalism based on propagators.

\item Spin glasses at non-zero magnetic field. Here the situation is worse: we lack both the lower and the
upper critical dimensions. At the naive upper critical dimensions (i.e. $D_U=6$) the trivial fixed is not infrared
stable and no infrared stable fixed point is known \cite{63}. The infrared stable fixed point could be a strong
coupling one, like in four-dimensional QCD, or a zero temperature fixed point; the situation calls for
further investigation: many different alternative scenarios are possible. 

\item Surface roughening in the framework or the KPZ equation \cite{69} for surface growth in $D+1$
dimension. (The model can also be mapped on the statistical properties of an equilibrium directed polymer in a random
medium.) Here the lower critical dimension is known (i.e. $D_L=0$). The problems come with the upper critical
dimension. From time to time arguments (albeit not completely convincing) are presented  \cite{70}  that hint at
$D_U= 4$, but this result seems to be at variance with numerical
simulations that suggest a higher value \cite{72}. A real space renormalization group computation suggests that the
upper critical dimension is infinite, in agreement with the numerically accurate (but not exact) formula for
the critical exponents $\eta$ controlling the width of the interface (i.e. $\eta =2/(2D+1)$). 
\item The DLA (direct limited aggregation) problem. This problem is rather difficult and only a few
attempts have been done to study the problem in large dimension \cite{73}. There is the very interesting suggestion
that the fractal dimension of the aggregate goes to $D-1$ for large dimensions, but not much has been done in this
direction. 
\end{itemize}

It is clear that each problem has its own peculiarities. Different approaches should be used. In all the
cases, with the exclusion of the last one, the replica method can be used. In spin glasses in the low
temperature phase the replica symmetry is spontaneously broken in the mean field approximation, while in
the other cases it remains exact (with maybe the exception of electron glasses). 
In the next sections I will discuss in more details a few case.

\section{A difficult simple case. The random magnetic field  Ising ferromagnet.}

In the nutshell, the random magnetic field  Ising ferromagnet is  a standard $\phi^4$ theory with a quenched magnetic field.
The probability of the field $\phi$ at given $h(x)$ is 
\be
P(\phi|h)\propto \exp(- \beta S(\phi|h))
\ee
with
\be
S(\phi|h)=\int dx\left(\frac12 (\partial \phi(x))^2 +\frac12 \tau  \phi(x)^2 +\frac14  \phi^4(x)-h(x)\phi(x)\ \right).
\ee
Of course we must specify the probability distribution of the magnetic field $h(x)$. We assume short-range correlations:
\be
P(h)=\exp \left(\frac{1}{2g}{\int dx h^2(x)}\right)\ , \ \ \ \ \ba{h(x)h(y)}=g \delta(x-y) \ ,
\ee
where the overline denotes the average over the $h$ distribution.

Going from positive $\tau$ to large negative $\tau$ we expect to find a phase transition: in the region of strongly negative $\tau$ the expectation value of $\phi$ becomes different from zero and the $Z_2$ symmetry of the model is spontaneously broken.
A detailed and convincing analysis shows that critical behavior remains the same also in the limit $\beta \to \infty$. A similar analysis show that near the critical point we have
\begin{equation}
\ba{\lan\phi(x)\phi(0)\ran_h}\approx \ba{\phi^*(x|h)\phi^*(0|h)} \ .
\end{equation}
where $\phi^*(x|h)=\lan \phi(x) \ran_h$ is the expectation value of the field $\phi(x)$ in presence of a field $h$.

Let us suppose that the solution of the following stochastic equation is unique
\be
{\delta S(\phi|h)\over \delta \phi(x)}= -\Delta \phi(x) +\tau \phi(x) +\phi(x)^3- h(x)=0\ . \label{SSE}
\ee
Let us call it $\phi(x|h)$. In the limit  $\beta \to \infty$  $\phi(x|h)=\phi^*(x|h)$. Under the previous assumptions it is possible to prove that
then
\be
\ba{\phi(x|h)\phi(0|h)}_D= \lan \phi(x)\phi(0)\ran _{D-2}\ .
\ee
where $\lan \cdot \ran_{D-2} $ is the statistical expectation value with weight $\exp(-S[\phi|0]/g)$ (i.e. at zero magnetic field) in two dimensions less \cite{65}.
This dimensional reduction follows from an hidden supersymmetry  of the stochastic differential equation.
Indeed if the solution of the stochastic differential equation is unique, after some transformations one arrives to a field theory that is invariant under the $O(D|2)$
supergroup. From this result dimensional reduction follows.
Unfortunately if the solution of the stochastic differential equation is not unique,
one can prove the beautiful and  useless identity:
\be
\lan \phi(x)\phi(0)\ran _{D-2}= \lan \phi(x)\phi(0)\ran_{SS}= \ba{\sum_\alpha I_\alpha \phi_\alpha(x|h)\phi_\alpha(0|h)} \, ,
\ee
where $\alpha$ label the solutions and $I_\alpha$ is the Morse index of the $\alpha^{th}$ solution and the subscript $SS$ denotes the supersymmetric results.

 If there are many solutions of eq. (\ref{SSE}),
$ \ba{\sum_\alpha I_\alpha \phi_\alpha(x|h)\phi_\alpha(0|h)}$ is very different from $\ba{\lan\phi(x)\phi(0)\ran_h}$. Indeed we expect that solutions with lower free energy have an higher weight and we should have
\be
\ba{\lan\phi(x)\phi(0)\ran_h}=\ba{\sum_\alpha I_\alpha \exp(-S[\phi_\alpha|h]/g) \phi_\alpha(x|h)\phi_\alpha(0|h) \over 
\ba{\sum_\alpha I_\alpha \exp(-S[\phi_\alpha|h]/g)}} \ .
\ee
 Unfortunately, when we approach the transition, $\tau$ (i.e. the {\em bare} mass squared) becomes negative (it is always negative in the continuum limit), $S(\phi|h)$ is no more convex and we expect many solutions in the critical region.

If we close our eyes on this problem, we can ask if dimensional reduction qualitatively works in three dimensions. The answer is no: the relation among the critical exponents of the random model in $D$ dimensions and the non-random model in $D-2$ dimensions, in particular
\be
\gamma_{RF}(D) =\gamma_P(D-2)\ ,
\ee
cannot be satisfied, where $\gamma_P(D-2)$ is the critical exponent of the pure system. The lower critical dimension is 2 for the random field problem (an exact result) and it is 1 for the non-random case. As far as $2\ne 1+2$, dimensional reduction fails badly.
We should notice that the existence of multiple solutions is a phenomenon that cannot be seen in perturbation theory. Only non-perturbative effects destroy dimensional reduction and formally the $6 -\eps$ expansion for random system coincides with the $4-\eps$ expansion for non-random system. A possible conjecture is
\begin{equation}
\gamma_{RF}(D) =\gamma_P(D-2) +O(\exp(-A/\eps))\ .
\end{equation}

Unfortunately it is not clear how to compute $\gamma_{RF}(D) -\gamma_P(D-2)$ . 
There are some suggestions (e.g. instantons), but no consensus. Moreover one could also take the view point that non-perturbative phenomena becomes important before we reach the transition \ and that the dimensional reduction formulae are not even approximatively valid \cite{DOT}.
The situation is quite confused and after nearly 30 years of work we are still far from a solution.

\section{Anderson localization: Schr\"odinger equations for non-interacting electrons}

Lt us  suppose that for some reasons we can neglect the interaction among electrons in a material. The properties of the electrons can be found by filling all the levels up to the Fermi level. At this end we have to find the solutions of the Schr\"odinger equations
\be
(-\Delta+V(x))\psi(x)=E \psi(x) \, ,  \ \ \ \ \  {\bf H} =-\Delta+V \, ,
\ee
where $V$ is the potential due to the remaining part of the systems. The interesting case is when the potential $V(x)$ is random (e.g. due the presence of impurities). Let us consider the simplest case: $V(x)$ uncorrelated from point to point with a Gaussian distribution:
\be
P(V)=\exp \left(\frac{1}{2g}{\int dx V^2(x)}\right) \ .
\ee

Roughly speaking we can divide the states (i.e. the eigenvalues of ${\bf H}$) into localized states and extended states. For localized states the spectrum is discrete and the wave function decays exponentially, e.g. as $\exp(-|x-x_0|/\xi)$; the typical value of $\xi$ is called the localization length.
For extended states, the spectrum is continuum: in a finite volume the wave function of an extended state spread everywhere and it is order of $O(V^{-1/2})$ in each point of the lattice.

It is possible to prove that for large negative $E$ there are localized states. 
In dimensions $D>2$ there is an threshold energy $E_L$ such that for $E<E_L$ all states are localized, while for $E>E_L$ all the states are extended. In dimensions $D\le2$ all states are localized.
If the electrons fill the levels up to the Fermi energy, the conductivity is related to
\be
G_2(x,\tilde E)=\ba{|G(x|\tilde E)|^2} \, ,\ \ \ G(x|\tilde E)=\lan x|{1 \over \tilde E-{\bf H}} |0\ran=
\sum_n {\psi_n (x) \psi_n (0) \over E-E_n}\ ,
\ee
where $\tilde E$ is complex ($\tilde E=E+i \eps$):
the real part ($E$) is the Fermi energy and the  imaginary part ($\eps$) is proportional to the frequency at which the conductivity is measured. Indeed one finds that the conductivity $\sigma$ is given by
\be
\sigma(\eps,E)\propto \eps^2 \left.{d \tilde G(k,\tilde E)\over d k^2} \right|_{k^2=0}\ ,
\ee
where $ \tilde G(k,\tilde E)$ is the Fourier transform of $ G(x,\tilde E)$.
In the following we will consider the
limit $\eps \to 0$. It is important to note that we cannot write the previous expressions for $G(x|\tilde E)$ and $G_2(x,\tilde E)$ directly in the limit $\eps=0$, because they would be undefined. Indeed 
\be
\lim_{\eps\to 0} \left(G(x| E+i\eps)-G(x| E-i\eps)\right)=2 \pi  i\sum_n \psi_n (x) \psi_n (0) \delta(E-E_n) \ .
\ee
This is similar to what is done in particle physics when we write $G(p)=1/ (p^2-m^2-i\eps)$.
Moreover in $G_2(x,\tilde E)$ there is a term that is proportional to \be
 \sum_n {|\psi_n (x)|^2 |\psi_n (0)|^2 \over (E-E_n)^2+\eps^2}\approx {1\over \eps} \sum_n |\psi_n (x)|^2 |\psi_n (0)|^2  \delta(E-E_n) \ .
\ee
It diverges (in distribution sense) when $\eps \to 0$. The divergence disappears in the infinite volume limit when the states are extended, but it is still present in the localized phase. In the localized phase the zero frequency conductivity vanishes and the correlation functions do diverge.

In the nutshell in the extended phase the correlation functions (e.g. $G_2(x,\tilde E)$) are long range, while in the localized phase the correlation functions are short range, but they diverge in the $\eps \to 0$ limit. Essentially we have two possibilities for small $\eps$ and $k$:
\be
\tilde G_2(k|\tilde E)= {A\over B \eps +k^2}\, ,\ \ \ \ \tilde G(k)={f(k)\over\eps}\ . \label{MIRA}
\ee
The first is realized in the extended phase and the second is realized in the localized phase. 

We would like to understand the behavior of the correlation length and of the conductivity near the localization transition, i.e. near $E_L$. At this end it is expedient to use a field theory representation.
We introduce $n$ real fields $\phi_a$ and $n$ real fields $\ba{\phi}_a$. \Cred{Eventually $n \to 0$, like in quenched QCD.} The theory in the $\eps \to 0$ limit has an Euclidean action  invariant under the \Cred{non-compact} group  $O(n,n)$. This group is defined as the set of transformation that leaves invariant the following quantity:
\Cblu{ 
\be
Q_2(x)\equiv\sum_{a=1,n} \left(\phi_a^2 -\ba{\phi}_a^2\right) \ .
\ee}
The explicit form of the action is 
\be
\int d[\phi] \exp(-S[\phi])\,  ,
\ee
where $S[\phi]$ is given by:
\be
i\int d^dx \left( \sum_{a=1,n}\left((\partial_\mu \phi_a(x))^2-(\partial_\mu \ba{\phi}_a(x))^2 -i \eps \left(\phi_a^2 \Cred{+}\ba{\phi}_a^2\right)\right)   +E Q_2(x)+g \left(Q_2(x)^2)\right) \right)
\ee
The factor $i$ in the action is crucial: the term $\eps \left(\phi_a^2 \Cred{+}\ba{\phi}_a^2\right)$ explicitly breaks the $O(n,n)$ symmetry, but is presence is necessary in order to make the functional integral convergent.

\Cblu{What happens to the $O(n,n)$ symmetry when $\eps \to 0$?}
This symmetry is broken in the non-localized phase and  in some sense it  becomes an exact symmetry in the localized phase. The peculiar behavior shown in eq. (\ref{MIRA}) is characteristic of  non-compact symmetry groups: the Green functions are infinite in the unbroken symmetry phase.
In this representation localization is a non-perturbative phenomenon (it arises as the effect of instantons). In order to compute the properties of the theory, we should study the critical behavior of a strongly interacting multi-instantons gas and this is not an easy task. Our inability to control the localization transition in this representation implies that the upper critical dimension in not yet known (may be it is infinite!).
At low dimensions things become simpler. We have seen that there are  no extended states in $D=2$ (it is a variation of a well known no-go theorem). There is an other field theory representation (a non-linear $\sigma$-model), where a  transition is present  in perturbation theory from the beginning and the exponents are know in $2+\eps$ dimensions.
In conclusion electron localization is an example of a theory where the lower critical dimension is know, but nothing sure is known about the upper critical dimensions.

\section{Spin glasses}

\Cblu{Many progresses have recently done in spin glasses: theory, experiments, simulations and even theorems!}
\Cblu{The simplest Ising spin glasses has the following Hamiltonian:}
	\be
	H_{J}(\sigma)=\frac12 \sum_{ik=1,N}J_{ik}\sigma_{i}\sigma_{k}
	\ee
	where the $\sigma$ are Ising spins (i.e. $\pm 1$), the $i$ and $k$  are {\em neighbors}. The $J$'s are random (e.g. $\pm 1$) and they are defined on the links of the lattice. If a magnetic field is present we must add to the Hamiltonian a term proportional to $\sum_i h\sigma_i$.
Let us present an alternative definition (in the continuum) that may be more appealing: they expose some of the symmetries of the problem. Let us consider a gauge group $G$, a gauge field $A_\mu(x)$ and a gauge transform $g(x)$.
In the Landau gauge we want  to minimize
\be
\int dx\Tr \left( A_\mu^g(x)\right)^2\equiv H_A(g).
\ee
We can consider (for a given gauge field) the probability distribution of $g$ given by
\be
\exp(-\beta H_A(g))\ .
\ee

This is a generalization of spin glasses, where $g$ and $A$ correspond to $\sigma$ and $J$. We obtain the  previous definition when the gauge group  $G$ is $Z_2$, we are on the lattice and we consider the strong coupling limit.
\Cblu{In many cases Gribov ambiguity tells us that $H_A(g)$ has many minima}, therefore we should not be surprised to known that $H_{J}(\sigma)$ has an exponentially large number of minima. We can ask many questions.
Which is the form of $H_A(g)$? How different are the minima? Which are the properties of set of minima (e.g. of their mutual distance?) Briefly which are the properties  of the (free) energy landscape?

This problem can be studied in the mean field approach where one can obtain exact results  \cite{MPV} (partially corroborated by theorems \cite{TALA}). It turns out that there is low temperature phase where the system has an exponentially large number of equilibrium states that are local minima of the free energy.
Each state has a very large (exponentially large) mean life and at equilibrium the system travels  from one state to an other state.

\Cblu{The system  has a very slow approach to equilibrium.} There is a fast dynamics inside the states and a very slow dynamics due to the barriers that separate the states (like punctuated equilibria in biology or in the landscape theory for strings). A peculiarity of this dynamics is the presence of violations of the fluctuation dissipation relations in slightly off-equilibrium dynamics. Numerical simulations give ample evidence of these new fluctuation-dissipation relations.   These new fluctuation-dissipation relations have been observed in some difficult experiments (notably by Herisson and Ocio in spin glasses \cite{38}).

May be the most characteristic phenomenon is the presence of two different magnetic susceptibilities that correspond to different paths in the temperature magnetic field plane. In the nutshell we have:
\begin{itemize}
\item The magnetic susceptibility that correspond to the response in one state ($\chi_{LR}$) that is measured by adding an infinitesimal magnetic field at fixed temperature.
\item The true equilibrium susceptibility, where one includes the effects of changing the equilibrium state when we change the magnetic field. It can be experimentally approximated by the field cooled magnetization $\chi_{FC}$.
\end{itemize}
The naive predictions of mean field theory and the experimental data agree in an impressive manner.

What happens in finite dimensions (we are particularly interested to the three dimensional case)? We would like to characterize the infrared (large scale) behavior in the low temperature phase and find in which dimensions the low temperature phase may exist. Numerical simulations are in good and detailed agreement with the predictions of a generalized replica approach in $D\ge 3$ (at least in zero magnentic field), while it is clear that in $D=2$ there is no transition (and the lower critical dimension is greater than 2).

By interpolating various exponent as function of the  dimensions one obtains the estimate $D_{L}=2.493(7)$ \cite{60}.  An explicit theoretical argument based on the evaluation of the interface energy gives the unexpected value $D_{U}=2.5$. Therefore we have a very good agreement among the theory and numerical simulations. 
 However the value $D_{L}=2.5$  is  strange. In the usual computation the lower critical dimension can
be related to the infrared divergence of some simple diagrams.  In scalar theories the spontaneous breaking of the $O(N)$ symmetry implies the presence Goldstone Bosons (the corresponding propagator being $1/k^2$).
The interaction is proportional to $k^2$ (soft pions theorems) and infrared divergences pile up in dimensions less or equal to 2. In spin glass we are unable, as we shall see later, to do such a simple computation. It is difficult to think which simple diagrams could diverge starting from 2.5
dimensions.  We would like to answer to the following questions.
Can we make a similar simple perturbative argument in spin glasses? Which is the low $k$ behavior of the propagators in D=3? We may get it from numerical simulations, but which are the theoretical predictions?

	Mean field theory was solved by replica approach. Before presenting the  field theory constructed as an expansion near the mean field theory \cite{MPV}, we must spend a few words in describing some of the unusual characteristic of the mean field. As usual we need to compute the average value of the logarithm of the partition function (i.e. the connected vacuum to vacuum diagrams): \be
	F\equiv -\ba {\ln (Z_J)} , \ee
	where $Z_J$ is the $J$ dependent partition function and the overline is the average over $J$. It turns out that
	it is simpler to compute first
	\be
	F_n=-{\log (\ba {Z_J^n})\over n}  \ .
	\ee
	 At the end we recover the value of $F$ using the formula
	 \be
	F =\lim_{n \to 0} F_n \ .
	\ee
	This approach leads to a strange mathematics: one introduces a  is $n \times  n$ matrix $Q_{a,b}$ ($a,b=1,\dots n$)  and the symmetry group is $S_n$ (the permutation group of $n$ elements); eventually $n$ goes to  $0$. In this way we construct a free energy $F[Q]$ and we obtain
	\be
	 F=F[Q^*]\, ,\ \ \ \ \mbox{where} \ \ \ \left. {\partial F\over \partial Q_{a,b}}\right|_{Q=Q^*}=0.
	 \ee

As a first approximation we suppose that the group of symmetry of the  system   ($S_n$ is the permutation groups of the $n$ replicas) is spontaneously broken to  it subgroup $S_{n/m} \otimes (S_m)^{n/m}$: the replica symmetry is spontaneously broken (it can be seen that spontaneous breaking of replica symmetry correspond to the existence of an infinite number of equilibrium states).
	 When $n\to 0$, $S_{n/m} \to S_0$ so that $S_0$ is a subset of  $S_0$ and the unbroken subgroup contains $S_0$. This fact implies the possibility of breaking the permutation group in an infinite hierarchical way. The permutation group $S_0$ is promoted to a continuous group. Indeed one can introduce infinitesimal permutation and write the Ward identities associated \cite{57,DDG}.
	  At the end of the day the matrix $Q$ is parametrized in terms of a function $q(z)$ defined on the interval $0-1$. One finally finds a free energy $F[q]$ (it a a functional of $q(z)$) that must be maximized (not minimized!) \cite{MPV,GUERRA}. Talagrand theorem \cite{TALA} tells us that this baroque construction  gives the exact result in the Sherrington Kirkpatrick model.
	
The perturbation around mean field theory can be formally constructed. There are no difficulties in the high temperature phase (the unbroken symmetry phase is conventional). We know the upper critical dimension (i.e. $D_U=6$) and we can
control the critical exponents in the $6-\eps$ expansion (three loops have been computed). The problems come in the low temperature phase where we do not  have a good control of the infrared behavior. 
The perturbative expansion has been constructed, but the evaluation of the results is difficult. As usually one starts by  computing the propagator at tree level:
\be
G_{a,b;c,d}(x)=\lan Q_{a,b}(x)Q_{c,d}(0)  \ran_c \, ,
\ee 
(we will denote by $ \tilde G_{a,b;c,d}(k)$ the Fourier transform of $G_{a,b;c,d}(x)$.

The propagator depends on four indices and in the replica broken phase it can parametrized in terms of functions of three variables $z_1,z_2,z_3$. The final form of the propagator can be obtained by solving six coupled integral equations \cite{57,DDG}. Depending on the choice of the indices ($a,b,c,d$) $\tilde G_{a,b;c,d}(k)$ has a behavior near $k=0$ of the type
\be
O(1),\ \ O(1/k^2), \ \ O(1/k^3),\ \ O(1/k^4)\ .
\ee
These long-range correlations (i.e. massless particles) are
related to the existence of Goldstone modes of a symmetry that is spontaneously broken in the low temperature
phase and some of them are consequences of the Ward identities \cite{57,DDG} associated to the infinitesimal permutations.
Moreover there is an infinite number of poles at $k^2=-m_i^2$, where the index $i$
goes up to infinity and the $m_i$'s accumulate toward zero \cite{57}. The one loop corrections can be written in an explicit way. If we proceed in a naive way we find at the end  $O(10^4)$- $O(10^6)$ different integrals and the computation is nearly unfeasible. In the infrared region there are very strong cancellation among the various terms (e.g. \cite{FP}) so that a very  careful evaluation is needed. It clear that one should regroup the various terms in a smart way, may using Ward identities. We have not been sufficiently smart.

The final effect of this complicated behavior is that we are not able do construct
the appropriate non-linear sigma model that should be used near the lower critical dimensions. The Ward
identities of the symmetry have been identified and they have been shown to be the responsible of many
important cancellations, but we are still far from writing the final effective theory \cite{57}. Maybe we are very near,
but we miss at least one crucial ingredient and we are stuck.

\section{The topological expansion: Fat Diagrams}
In this section I would only mention a new idea that has bee put forward by Efetov \cite{E,PSL,VS} about twenty years ago. It looks very promising; however not too much progress has ben done because the needed computations are feasible, but technical difficult.

This approach can be defined only for a theory on the lattice. The typical Hamiltonian is of the form:
\be
H=\sum_{i,k} H_{i,k}(\phi_i,\phi_k)\, ,
\ee
where the sum is done over the nearest neighbor points, the Hamiltonian may be non-translational invariant (has happens in a random system) and the field $\phi_i$ may be a vector with many components.
Our  aim is to find out the relevant degrees of freedom in the continuum limit, i.e. near the phase transition where the the correlation length becomes infinite. There are many ingredients that enter in the approach: a)
the Bethe version of mean field theory that becomes exact on the  Bethe lattices; b) the control of the one dimensional version of the model; c)  the high temperature expansion in lattice theories.

Let me try to summarize the basic ideas. In mean field theory, one neglect the loops and one looks for a self-consistent solution of the equation of motion. This corresponds to neglect correlations and to impose self-consistent boundary conditions. Mean field theory becomes exact in the case where the lattice is locally loopless (i.e. it has the topology of a tree): loops are present, but they have a length that goes to infinity with the volume (this happens on Bethe lattices). The presence of large loops accounts for the need of imposing self consistent conditions for the vacuum expectation values. 

In the topological expansion one starts from the mean field theory, where the lattice is locally a tree. In order to describe the approach it is convenient to recall the particle representation in field theory.
In $D$ dimensions the free propagator (i.e. $1/(k^2+m^2)$ in momentum space) can be written in configurations space as:
\be
G(x)=\int ds \cN(x,s) \exp (-m^2s) \label{TREE} \, ,
\ee
where $\cN(x,s)\propto s^{-D/2} \exp(-x^2/s)$ is the probability of a path of length  (internal time)  $s$ going from $0$ to $x$. This representation is well known in the continuum, but it can be extended also to the lattice.

A one  loop contribution for the propagator in a $\phi^3$ theory can be written in a similar way as   
\bea
\int d^Dy \, d^D z \,G(x-y)G(x-z)^2G(z)=\int d^Dy \, d^D z \int ds_1  \, ds_2  \, ds_3  \, ds_4  \label{LOOP} \\
\exp( -m^2(s_1+s_2+s_3+s_4))\cN(x-y,s_1)\cN(y-z,s_2)\cN(y-z,s_3)\cN(z,s_4)\ .
\eea

In the {\sl fat} diagram approach we will derive similar formulas in a different context:  we have to consider a theory on the lattice in one dimensions in a fluctuating background and at the end, in eq. (\ref{TREE}),
we will substitute  $\exp (-m^2s)$ with the correlation function on an one-dimensional model.
We substitute  in eq. (\ref{LOOP})
 $\exp( -m^2(s_1+s_2+s_3+s_4))$ with the correlation function on an one-dimensional model, with the topology of the corresponding Feynmann graph.

Let us see how to proceed in details. If we neglect loops we would naively write the following equation for the probability distribution of the field $\phi$:
\be
P(\phi)\propto \prod_{k=1,z}\left(\int d\phi_k P(\phi_k) \exp(-H(\phi,\phi_k)\right)= \left(\int d\phi' P(\phi') \exp(-H(\phi,\phi')\right)^z \ ,
\ee
where $z$ in the coordination of the lattice (e.g. $z=2D$ in an hypercubic lattice). However the previous formula is not correct because the fields in two nearest neighbour points are correlated. The correct results for a loopless system is given by the Bethe approximation.
We call $B(\phi)$ the probability distribution the field with $z-1$ neighbours (we remove one spin and we consider the probabability distribution of the spins around this cavity). After some work get the equations:
\be
B(\phi)\propto\left(\int d\phi' B(\phi') \exp(-H(\phi,\phi')\right)^{z-1}
\, , \ \ \ \ \ \
P(\phi) \propto\left(\int d\phi' B(\phi') \exp(-H(\phi,\phi')\right)^z \ .
\ee

Let us compute the correlations in this framework. Let us suppose that there is an unique path from point $x$ to $y$ of lentgh $L$.
We can compute the probability
 \be
P_c^{(L)}(\phi(x),\phi(y))\equiv P(\phi(x),\phi(y)) -P(\phi(x))P(\phi(y)) \ .
\ee
 by solving the one dimensional problem where each site interacts with $z-2$ sites outside the unique one dimensional path. Let us assume that $P_c(\phi(x),\phi(y))$ is small, so that if there are many path they do not interfere.
We finally get the zero loop contribution:
\be
P_c^0(\phi(x),\phi(y))=\sum_L\cN(L,x-y)P_c^{(L)}(\phi(x),\phi(y)) \ .
\ee
\nuova

The effects of the loops can be included in a systematic way, by considering the contribution of lattice region with high genus. In this way one obtains new objects (that we  can call them {\sl fat} diagrams): they look like usual diagrams, but they contain the resummation of non-perturbative effects in one dimensions. Their computation  is rather complex, but it seems to be feasible. At the end of the day we get a genus expansion:
\be
P_c(\phi(x),\phi(y))=\sum_{g=0,\infty} P_c^g(\phi(x),\phi(y))\, ,
\ee
where $P_c^g(\phi(x),\phi(y)$ can be computed on a lattice of genus $g$.
\section{Conclusions}

We have seen that there are many simple disorder systems where we do not have a good control of physics (e.g. we do know the behavior near the upper critical dimension, if any, and near the lower critical dimension):
\begin{itemize}
	\item Equilibrium systems, e.g.  ferromagnets with a random field, localized electrons, directed polymers, spin glasses (both with and without a field) .
	\item Non-equilibrium systems, e.g. surface growth (KPZ equation), diffusion limit aggregation (DLA).
\end{itemize}

In particular we do not control the large scale behavior of spin glasses in the broken symmetry phase.
Many of these problems stand for twenty years or more.
We badly need new ideas, new tools, techniques and also smart people trying to solve them.
Fat diagrams is a promising possibility. I hope that they will fulfill the promises!

\end{document}